\documentclass[10pt,a4paper,english]{article}
\usepackage[english]{babel}
\usepackage[utf8]{inputenc}
\usepackage{amsmath}
\usepackage{amsfonts}
\usepackage{amssymb}
\usepackage{graphicx}
\usepackage{slashbox}

\newcommand{\be}{\begin{equation}}
\newcommand{\ee}{\end{equation}}

\newcommand{\aaa}{\alpha}
\newcommand{\piinsc}{\pi_0^\alpha}
\newcommand{\taa}{\tau^\alpha}
\newcommand{\toa}{\tau^\alpha_0}
\newcommand{\saa}{\sigma^\alpha_0}

\begin{document}
\title{\begin{Large}\begin{flushleft} \textit{Short Communication}\\ \textbf{Three quantitative predictions based on past regularities about voter turnout at the French 2009 European election}\end{flushleft}\end{Large}}
\author{Christian Borghesi\thanks{Electronic address: christian.borghesi@cea.fr}\\
\textit{Service de Physique de l’\'Etat Condens\'e, Orme des Merisiers,}\\ \textit{CEA Saclay, Gif sur Yvette Cedex, 91191, France.}}
\date {(May 27, 2009)}
\maketitle
\begin{abstract}
\normalsize
The previous twelve turnout rates of French national elections by municipality show regularities. These regularities do not depend on the national turnout level, nor on the nature of the election. Based on past statistical regularities we make three predictions. The first one deals with the standard deviation of the turnout rate by municipality. The second one refers to the continuity in time of the heterogeneity of turnout rates in the vicinity of a municipality. The last one is about the correlation between the heterogeneity of turnout rates in the vicinity of a municipality and the population in its surroundings. Details, explanations and discussions will be given in forthcoming papers.
\end{abstract}
\vspace{0.5cm}
Most of the empirical electoral studies made by physicists deal with proportional voting from multiple choice lists, and investigate the distribution of votes ; as in Brazil~\cite{costa_filho_scaling_vot, costa_filho_bresil_el2, lyra_bresil_el, bernardes_bresil_el}, in Brazil and India~\cite{gonzalez_bresil_inde_el}, in India and Canada~\cite{hit_is_born}, in Mexico~\cite{baez_mexiq_el, morales_mexiq_el}, in Indonesia~\cite{situngkir_indonesie_el}. A universal behavior is seen in~\cite{fortunato_universality}. Statistical results of elections for city mayor, focused on few candidates, are studied in~\cite{araripe_plurality}, typology of Russian elections in~\cite{sadovsky_russie_el}, a correlation between electoral results and party members in Germany in~\cite{schneider_impact}, and statistics of votes per cabin for three Mexican elections in~\cite{hernandez_bvot_mexique}.

In this paper, we are interested in a probably simpler phenomenon: the participation in elections. Table~\ref{tabst} shows the previous twelve French national elections for which we know the turnout rate for each of the approximately $36200$ municipalities (called ``communes'' in French)\footnote{Data are from the \textit{bureau des \'elections et des \'etudes politiques} of the French Home Office, kindly given by Brigitte Hazart, who I would like to especially thank for the great work she did.}. It is worth to stress that for each one of these elections (first or second presidential round, referendum, European Parliament), the choice list is the same for the whole country\footnote{This is strictly true in all elections considered, excepted in the 2004 and the future 2009 European Parliament elections, in which there are seven different choice lists over metropolitan France. In the latter case, the main political parties propose candidates in each of the seven regions. Nevertheless, we assume that, regarding voter turnout, it is not a great mistake to assimilate these seven lists to a single one.}. This investigation is restricted to metropolitan France.

The three predictions are produced by statistical regularities observed over the previous twelve elections, while global turnout rates stretch from $31\%$ up to $85\%$. Moreover, these elections are not of the same nature. A standard deviation (sigma) is measured, and a prediction is made with an arbitrary \textit{two sigma} error bar. Note that for each one of the three regularities, data do not seem to show any global shift during the studied fifteen year period.

\section{Standard deviation of $\taa$}

Let $\aaa$ be a municipality, with $N^\aaa$ registered voters, of whom $N_+^\aaa$ take part in the election, and the remaining $N_-^\aaa$ do not ($N^\aaa = N_+^\aaa + N_-^\aaa$). In order to compare in an easier way elections with different turnout levels, let us define the unbounded value for one municipality $\aaa$:
\be \label{etaa} \taa = ln\big(\:\frac{N_+^\aaa}{N_-^\aaa}\:\big)~.\ee
$\taa>0$ means that the usual turnout rate $\frac{N_+^\aaa}{N^\aaa}$ is greater than $50\%$. To prevent $\taa$ from diverging, we modify real values by $\frac{1}{2}$ when necessary. (If $N_-^\aaa=0$, then $N_-^\aaa\rightarrow\frac{1}{2}$, $N_+^\aaa\rightarrow N^\aaa-\frac{1}{2}$ ; and if $N_+^\aaa=0$, then $N_+^\aaa\rightarrow\frac{1}{2}$, $N_-^\aaa\rightarrow N^\aaa-\frac{1}{2}$.)

\begin{figure}[t]
 \begin{minipage}[c]{0.46\linewidth}
  \includegraphics[width=8cm, height=4cm]{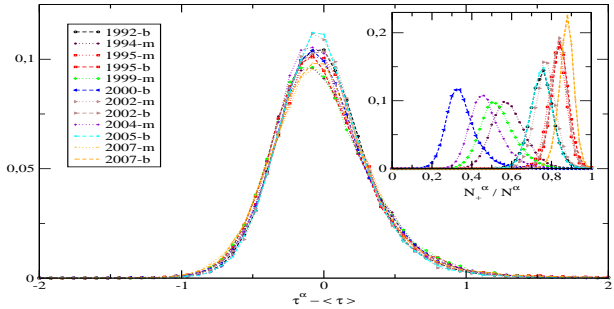}
 \end{minipage} \hfill
 \begin{minipage}[c]{0.46\linewidth}
  \caption{\small Histograms of $\taa - \langle\tau\rangle$, for the twelve elections. $\langle\tau\rangle$ means the average value of $\taa$ over the approximately $36200$ municipalities $\aaa$ of metropolitan France. The letter $b$ (or $m$) added to the election year indicates that the choice list is binary (or multiple). For more information about elections, see Tab.~\ref{tabst}. The inset shows the histograms of the turnout rate $\frac{N_+^\aaa}{N^\aaa}$ per municipality.}
 \label{fhisto-log-abst}
 \end{minipage}
\end{figure}

Fig.~\ref{fhisto-log-abst} shows centered histograms of $\taa$ for the twelve elections studied. The centered distribution ($\taa-\langle\tau\rangle$) seems to be similar for the twelve different elections. It is worth to stress that the similarity of the $\taa$ distribution cannot be explained by a simple binomial distribution, where $N^\aaa_+$ is obtained by the successes of $N^\aaa$ independent events with probability $p$. (According to the binomial distribution hypothesis, the probability $p$ is the same for all the municipalities, and its value, which depends on the considered election, is equal to the global turnout rate given in Tab.~\ref{tabst}.) Note that other unbounded values, e.g. $-erfc^{(-1)}(\frac{2N^\aaa_+}{N^\aaa})$, can also lead to similar centered distributions for the twelve different elections. We choose the definition given in Eq.~(\ref{etaa}) because of its simplicity.

Here, we are just interested in the standard deviation of the $\taa$ distribution, which can easily be measured, and thus predicted. Tab.~\ref{tstat-log-abst} gives the standard deviation of $\taa$ over all the municipalities $\aaa$, for each election.

\begin{table}[t]
\begin{tabular}{|c|c|c|c|c|c|c|c|c|c|c|c|}
\hline
92-b & 94-m & 95-m & 95-b & 99-m & 00-b & 02-m & 02-b & 04-m & 05-b & 07-m & 07-b\\
\hline
0.355 & 0.398 & 0.375 & 0.398 & 0.392 & 0.377 & 0.347 & 0.367 & 0.366 & 0.351 & 0.396 & 0.394\\
\hline
\end{tabular}
\caption{\small Standard deviation of $\taa$ over all the municipalities. The mean value and the standard deviation over the twelve elections are respectively equal to $0.376$ and $0.019$.}
\label{tstat-log-abst}
\end{table}

\begin{table}[h!]
\hspace{0.5cm}Thus, the predicted standard deviation of $\taa$ over all the municipalities becomes:
\begin{center}
\begin{tabular}{|c|c|c|}
\cline{2-3}
\multicolumn{1}{c|}{}
 & Previous measures & Expected measure\\
\hline
Standard deviation of $\taa$ & $0.376 \pm 0.019$ & $[0.338 ; 0.414]$\\
\hline
\end{tabular}\\
\begin{flushleft}(In the above table, \textit{Previous measures} is written as $(mean\pm~standard\;deviation)$, like in the next two similar ones.)\end{flushleft}
\end{center}
\end{table}

It can be interesting to compare the turnout rate of a central municipality $\aaa$ to its environment, defined as the $n_v$ municipalities $\beta$ in the vicinity of $\aaa$~\cite{thesis, preparation}. This environment, i.e. the set of the $n_v$ different $\tau^\beta$, can be characterized by its mean value, and also by its standard deviation. In order to take into account more properly the standard deviation of the environment, we define it as a constant number $n_v$ (the same for each central municipality) of the nearest neighbor municipalities of a central municipality. It appears that generally, this standard deviation, called here $\saa$, i.e. the heterogeneity of the environment of a central municipality $\aaa$, remains relatively stable for each election. More precisely, the correlation over all the central municipalities $\aaa$ between $\saa$ at two different elections, is relatively high (around $0.6$), and does not fluctuate a lot for different couples of elections.
\section{Correlation of $\mathbf{\saa}$ at different elections}

Consider the $n_v=16$ nearest neighbor municipalities of a central municipality $\aaa$. Note that the location of a municipality is reduced to the location of its town hall\footnote{\textit{Le Repertoire G\'eographique des Communes} by \textit{l'Institut G\'eographique National} provides latitude and longitude coordinates of the town hall of each municipality for 1999.\\http://professionnels.ign.fr/ficheProduitCMS.do?idDoc=5323862}. These $n_v$ municipalities have an estimated standard deviation of their turnout rates, written as:
\be \label{esaa} \saa = \sqrt{\frac{1}{n_v-1}\sum_{\beta}(\tau^\beta-\toa)^2}~,\ee
where $\beta$ is one of the $n_v=16$ municipalities, and $\toa=\frac{1}{n_v}\sum_\beta\tau^\beta$.

The correlation of $\saa$ for all the $\aaa$ municipalities at two different elections, at time $t_i$ and $t_j$, is written as $C_{t_i,t_j}(\sigma_0)$. ($C_{t_i,t_j}(\sigma_0)=\frac{\langle \saa(t_i)\saa(t_j) \rangle - \langle \saa(t_i)\rangle \langle \saa(t_j)\rangle}{\sqrt{\langle[\saa(t_i)-\langle \saa(t_i)\rangle]^2\rangle} \sqrt{\langle[\saa(t_j)-\langle \saa(t_j)\rangle]^2\rangle}}$, where $\saa(t_i)$ and $\saa(t_j)$ respectively mean $\saa$ at times $t_i$ and $t_j$, and $\langle ...\rangle$ means the average value over the $\aaa$ municipalities.) Table~\ref{ttempo-saa-abst} gives $C_{t_i,t_j}(\sigma_0)$ for every couple of different elections $t_i\neq t_j$.

\begin{table}[h!]
\begin{tabular}{|c|c|c|c|c|c|c|c|c|c|c|c|}
\hline
\backslashbox{$t_i$}{$t_j$} & 94-m & 95-m & 95-b & 99-m & 00-b & 02-m & 02-b & 04-m & 05-b & 07-m & 07-b\\
\hline
1992-b & 0.636 & 0.602 & 0.600 & 0.597 & 0.559 & 0.501 & 0.521 & 0.543 & 0.515 & 0.484 & 0.520\\ 
1994-m & & 0.609 & 0.595 & 0.686 & 0.607 & 0.524 & 0.509 & 0.603 & 0.535 & 0.463 & 0.505\\
1995-m & & & 0.732 & 0.607 & 0.555 & 0.551 & 0.549 & 0.554 & 0.531 & 0.507 & 0.535\\
1995-b & & & & 0.605 & 0.547 & 0.537 & 0.554 & 0.556 & 0.524 & 0.500 & 0.536\\
1999-m & & & & & 0.697 & 0.589 & 0.570 & 0.700 & 0.592 & 0.510 & 0.551\\ 
2000-b & & & & & & 0.552 & 0.546 & 0.662 & 0.542 & 0.485 & 0.532\\
2002-m & & & & & & & 0.694 & 0.581 & 0.557 & 0.499 & 0.535\\
2002-b & & & & & & & & 0.566 & 0.572 & 0.530 & 0.574\\
2004-m & & & & & & & & & 0.642 & 0.535 & 0.575\\
2005-b & & & & & & & & & & 0.567 & 0.600\\
2007-m & & & & & & & & & & & 0.704\\
\hline
\end{tabular}
\caption{\small $C_{t_i,t_j}(\sigma_0)$ for all the couples of different elections. The mean value and the standard deviation are respectively equal to $0.567$ and $0.058$.}
\label{ttempo-saa-abst}
\end{table}

\begin{table}[h!]
\hspace{0.5cm}Thus, the predicted $\frac{1}{12}\sum_{i=1}^{12} C_{t_i,t_j}(\sigma_0)$, where $i$ is one of the previous twelve elections and $j$ the next European Parliament election, becomes:
\begin{center}
\begin{tabular}{|c|c|c|}
\cline{2-3}
\multicolumn{1}{c|}{}
& Previous measures & Expected measure\\
\hline
$C_{t_i,t_j}(\sigma_0)$ & $0.567 \pm 0.058$ & $[0.451 ; 0.683]$\\
\hline
\end{tabular}
\end{center}
\end{table}

The dispersion of turnout rates in the surroundings of a municipality shows a constant behavior in time. We will now see its connection with the population in the vicinity of the municipality.
\section{Correlation between $\mathbf{\saa}$ and $\mathbf{\piinsc}$}

\begin{figure}[t]
 \begin{minipage}[c]{0.46\linewidth}
  \includegraphics[width=8cm, height=4cm]{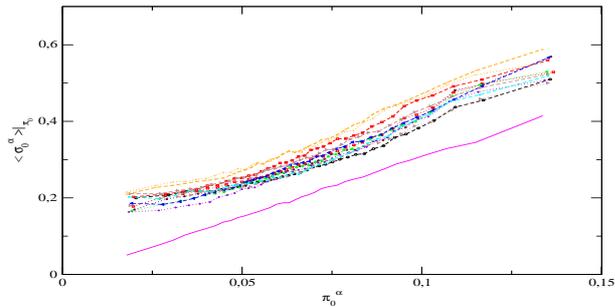}
 \end{minipage} \hfill
 \begin{minipage}[c]{0.46\linewidth}
  \caption{\small Mean value of $\saa$ inside $\piinsc$ intervals, for the twelve elections. There are $36$ intervals, and each one contains roughly $1000$ $\saa$. Legends are the same as in Fig.~\ref{fhisto-log-abst}. In magenta, data derive from a binomial simulation, where $N^\aaa_+$ is obtained by the successes of $N^\aaa$ independent events with the same probability $p$ for each municipality $\aaa$. In this simulation, $N^\aaa$ and $p$ are real values of the 2007-b election.}
 \label{fsaa-piinsc-abst}
 \end{minipage}
\end{figure}

Finite size effects are partially included in $\saa$, and are due to the finite size of the population of the $n_v$ municipalities that define $\saa$. These finite size effects can be taken into account by the value $\piinsc$, written as:
\be \label{epiinsc} \piinsc = \sqrt{\frac{1}{n_v} \sum_{\beta} \frac{1}{N^\beta}}~,\ee
where $\beta$ means one of the $n_v=16$ nearest neighbor municipalities of a central municipality $\aaa$. According to Eq.~(\ref{esaa}), $N^\aaa_+$ obtained by the successes of $N^\aaa$ independent events with probability $p$ (i.e. a binomial distribution), provides $\saa$ proportional to $\piinsc$ (see Fig.~\ref{fsaa-piinsc-abst}).

Fig.~\ref{fsaa-piinsc-abst} shows $\saa$ as a function of $\piinsc$ for each election. There is a clear connection between $\saa$ and $\piinsc$, that can be evidenced by their correlation. Table~\ref{tsaa-piinsc-abst} gives for each election the correlation between $\saa$ and $\piinsc$ over all the central municipalities $\aaa$.

\begin{table}[h!]
\begin{tabular}{|c|c|c|c|c|c|c|c|c|c|c|c|}
\hline
92-b & 94-m & 95-m & 95-b & 99-m & 00-b & 02-m & 02-b & 04-m & 05-b & 07-m & 07-b\\
\hline
0.636 & 0.633 & 0.665 & 0.662 & 0.661 & 0.677 & 0.615 & 0.635 & 0.679 & 0.609 & 0.610 & 0.663\\
\hline
\end{tabular}
\caption{\small Correlation between $\saa$ and $\piinsc$ over all the central municipalities $\aaa$, and for the twelve elections. The mean value and the standard deviation are respectively equal to $0.645$ and $0.026$.}
\label{tsaa-piinsc-abst}
\end{table}

\begin{table}[h!]
\hspace{0.5cm}Thus, the predicted correlation between $\saa$ and $\piinsc$ becomes:
\begin{center}
\begin{tabular}{|c|c|c|}
\cline{2-3}
\multicolumn{1}{c|}{}
 & Previous measures & Expected measure\\
\hline
Correlation between $\saa$ and $\piinsc$ & $0.645 \pm 0.026$ & $[0.593 ; 0.697]$\\
\hline
\end{tabular}
\end{center}
\end{table}

To conclude, we have been interested in a relatively simple phenomenon: the participation in elections, in France. We have observed three regularities of turnout rate per municipality (or ``commune'' of France). (For each of the twelve elections studied, there is the same choice list for metropolitan France.) The first one is the standard deviation over the approximately $36200$ municipalities of $\taa$ (an other way to write the turnout rate). The second one involves the constant behavior in time of $\saa$ (the standard deviation of turnout rates in the vicinity of a municipality). The third one is about the connection of $\saa$ with $\piinsc$ (the population distribution in the surroundings of a municipality). These three regularities do not depend on the turnout level, nor on the nature of the election. These facts incite us to make three predictions for the next 2009 European Parliament election, within an arbitrary \textit{two sigma} error bar.

We are convinced that science, fortunately, is not limited to observe empirical regularities. Nevertheless, knowing regularities allows to make irregularities significant, when they occur (as it is shown in~\cite{thesis}). But, what matters most is what we try to explain and to use (as we try to do in~\cite{thesis, preparation}) some of the encountered regularities, in order to pursue theoretical investigation.

\vspace{0.5cm}
{\bf Acknowledgments\\}
The brilliant idea to study electoral data in connection with their location is entirely due to Jean-Philippe Bouchaud, and I would like to thank him for this. A great thank to Ivan Dornic and Lionel Tabourier for their help and for their passionate discussions ; to Diana Garc\'{i}a L\'{o}pez and Marie-Alice Gouin for their help with the translation.

\vspace{0.5cm}
\renewcommand{\theequation}{A-\arabic{equation}}
\setcounter{equation}{0}  
\renewcommand{\thefigure}{A-\arabic{figure}}
\setcounter{figure}{0}
\renewcommand{\thetable}{A-\arabic{table}}
\setcounter{table}{0}
\appendix
\begin{large}
\textbf{\appendixname{ A : Elections studied for the voter turnout}}
\end{large}
\label{annexe-etudiees}
\begin{table}[h!]
\begin{tabular}{|c|c|c|c|c|c}
\hline
Year & List & Election & Registered voters & Turnout rate\\
\hline
1992 & binary & referendum, Maastricht Treaty & 36.6 $10^6$ & 0.713\\
1994 & multiple & European Parliament election & 37.6 $10^6$ & 0.539\\
1995 & multiple & presidential election, first round & 38.4 $10^6$ & 0.795\\
1995 & binary & presidential election, second round & 38.4 $10^6$ & 0.805\\
1999 & multiple & European Parliament election & 38.4 $10^6$ & 0.478\\
2000 & binary & referendum, five year presidential period & 38.2 $10^6$ & 0.308\\
2002 & multiple & presidential election, first round & 39.2 $10^6$ & 0.729\\
2002 & binary & presidential election, second round & 39.2 $10^6$ & 0.810\\
2004 & multiple & European Parliament election & 39.9 $10^6$ & 0.434\\
2005 & binary & referendum, European constitutional Treaty& 39.7 $10^6$ & 0.711\\
2007 & multiple & presidential election, first round & 41.9 $10^6$ & 0.854\\
2007 & binary & presidential election, second round & 41.9 $10^6$ & 0.853\\
\hline
\end{tabular}
\caption{\small Elections studied for their voter turnout. The choice list can be binary or multiple, and is uniform over the country. The data are limited to metropolitan France.}
\label{tabst}
\end{table}

\end{document}